\documentclass[aps,twocolumn,preprintnumbers,groupedaddress,superscriptaddress,floatfix,
tightenlines,reprint,nofootinbib,longbibliography]{revtex4-1}
\usepackage{mathrsfs}
\usepackage{natbib}
\usepackage{verbatim}
\usepackage{enumerate}
\usepackage{graphicx,bbm,amsbsy,amsfonts,amssymb,amsmath}
\usepackage{bm}
\usepackage{hyperref}
\usepackage{booktabs}
\usepackage{gensymb}
\usepackage{subfigure}
\usepackage{xcolor}
\usepackage{footmisc}
\usepackage{cancel}
\usepackage{blindtext}
\usepackage{ytableau}

\def\2{{(2)}}
\def\1{{(1)}}
\def\0{{(0)}}
\def\m1{{(-1)}}

\let\oldAA\AA
\renewcommand{\AA}{\text{\normalfont\oldAA}}

\hypersetup{
    colorlinks=true,       
    linkcolor=red,          
    citecolor=blue,        
    filecolor=magenta,      
    urlcolor=blue           
}

\newcommand{\be}{\begin{equation}}
\newcommand{\ee}{\end{equation}}
\newcommand{\ea}{\end{eqnarray}}
\newcommand{\ba}{\begin{eqnarray}}
\begin{document}


\preprint{UMN-TH-4516/25}
\preprint{FTPI-MINN-25-17}

\title{An Extra-Dimensional Axion in a 5D Warped Orbifold GUT}

\author{Gongjun Choi}
\email{choi0988@umn.edu}
\affiliation{William I. Fine Theoretical Physics Institute, School of Physics and Astronomy,\\
University of Minnesota, Minneapolis, Minnesota 55455, USA}

\author{Tony Gherghetta}
\email{tgher@umn.edu}
\affiliation{School of Physics and Astronomy, University of Minnesota, Minneapolis, Minnesota 55455, USA}

\begin{abstract}
 
We study the QCD axion arising from the 5th component of a bulk $U(1)$ gauge field in 
a five-dimensional warped grand unified theory, and determine the viable range of the axion decay constant $f_a$. Unlike flat extra dimensions, where gauge couplings run quickly above the Kaluza--Klein (KK) scale, the logarithmic running in warped geometries permits substantially smaller $f_a$ while preserving perturbative gauge coupling unification. However, bulk tree-level contributions to the gauge coupling---interpreted holographically as CFT renormalization---place a lower bound on $f_a$. We find that the conventional QCD axion window 
$10^{9}\,\mathrm{GeV} \lesssim f_a \lesssim 10^{12}\,\mathrm{GeV}$ is readily compatible without losing perturbativity, provided the AdS curvature is near the Planck scale. Thus, the 5D warped orbifold GUT naturally accommodates a high-quality QCD axion in a grand unified theory that provides an effective description of string-theoretic warped flux compactifications, admitting complementary geometric and holographic descriptions of the axion.
\end{abstract}


\date{\today}
\maketitle


\section{Introduction}
Among the enduring mysteries of the Standard Model (SM) lies the remarkable absence of CP violation in the strong interaction. Quantum chromodynamics (QCD) permits a topological term proportional to the angle $\bar{\theta}$, which, on natural grounds, would be expected to take a value of order unity. Yet experimental limits confine $\bar{\theta} \lesssim 10^{-10}$~\cite{Baker:2006ts}, giving rise to the \emph{strong CP problem}. A particularly elegant solution is offered by the QCD axion---a pseudo--Nambu--Goldstone boson (pNGB) emerging from the spontaneous breaking of an anomalous global $U(1)_{\rm PQ}$ symmetry---whose low-energy dynamics dynamically relax $\bar{\theta}$ to zero~\cite{Peccei:1977hh,Peccei:1977ur,Weinberg:1977ma,Wilczek:1977pj}. Intriguingly, the same field that resolves the strong CP problem can also serve as a dark matter candidate: axions produced non-thermally in the early universe through the misalignment mechanism naturally behave as cold dark matter, positioning the QCD axion as one of the most theoretically compelling extensions of the SM~\cite{Preskill:1982cy,Abbott:1982af,Dine:1982ah}.

Beyond the conventional Peccei–Quinn (PQ) construction, QCD axions can also emerge as zero modes of higher-form gauge fields in string-theoretic or higher-dimensional frameworks, e.g. Ramond–Ramond $p$-form gauge fields or Kalb-Ramond two-form gauge field~\cite{Witten:1984dg,Choi:1985je,Barr:1985hk,Choi:2003wr,Dvali:2005an,Svrcek:2006yi,Arvanitaki:2009fg}. These so-called extra-dimensional axions are often regarded as theoretically more robust than those originating from ad hoc global symmetries, since their interactions are dictated by underlying gauge invariance and genuine higher-form symmetries. This feature renders them intrinsically protected from potentially dangerous PQ-violating operators induced by quantum-gravitational effects~\cite{Burgess:2023ifd,Choi:2023gin,Craig:2024dnl,Reece:2025thc}. Unlike in conventional PQ models, where the axion decay constant $f_{a}$ is fixed by the scale of spontaneous $U(1)_{\rm PQ}$ breaking, the decay constant of an extra-dimensional axion is set by the higher-dimensional gauge coupling together with the geometry of the compact space—such as its size, shape or volume.

If the parameters determining $f_a$ in extra-dimensional axion frameworks are subject to non-trivial theoretical or phenomenological constraints, these relations can collectively yield meaningful bounds on the axion decay constant itself. A concrete illustration arises when the one-form extra-dimensional axion is embedded within a five-dimensional (5D) flat orbifold grand unified theory (GUT). As recently discussed in~\cite{Benabou:2025kgx}, ensuring the perturbativity of the unified gauge coupling requires the Kaluza--Klein (KK) threshold scale $m_{\rm KK}$ to remain above the GUT scale $M_{\rm GUT}$, since the gauge coupling exhibits a power-law running---linear in energy---beyond $m_{\rm KK}$~\cite{Dienes:1998vh,Dienes:1998vg}. This consideration translates into a non-trivial lower bound on the decay constant, $f_a \gtrsim 10^{14}\,{\rm GeV}$. A natural question then arises: is this constraint universal among extra-dimensional axion models, or does it rely on the specific geometry of the compact space? In particular, how might the situation change if the extra dimension is warped, as in the Randall--Sundrum (RS) construction~\cite{Randall:1999ee}?

Motivated by the observation that, for gauge fields propagating in a slice of AdS$_5$, the 4D effective gauge coupling continues to run only logarithmically even above the KK threshold~\cite{Pomarol:2000hp,Randall:2001gc,Randall:2001gb,Goldberger:2002cz,Goldberger:2002hb,Agashe:2002bx,Choi:2002wx,Contino:2002kc,Arkani-Hamed:2000ijo}, we investigate how small the axion decay constant $f_a$ can naturally be for an extra-dimensional one-form QCD axion in a 5D warped orbifold GUT. We show that maintaining perturbativity of the unified gauge coupling $g_{\rm GUT}$ at the unification scale $M_{\rm GUT}$ can be compatible with axion decay constants in the conventional QCD axion window, $10^{9}~{\rm GeV} \lesssim f_a \lesssim 10^{12}~{\rm GeV}$, provided the hierarchy between the AdS curvature scale $k$ and the compactification scale $R^{-1}$ is mild. Although we work within a supersymmetric 5D warped orbifold GUT, our setup may equally be viewed as a low-energy effective description of warped string compactifications in which axions descend from higher-dimensional gauge fields.

In Sec.~\ref{sec:5DGUT}, we investigate the renormalization group (RG) evolution of the four-dimensional (4D) effective gauge couplings when the minimal supersymmetric Standard Model (MSSM) gauge multiplets propagate in a slice of AdS$_5$, emphasizing the key feature that the running remains logarithmic even for momenta $q>m_{\rm KK}$, which controls the viable range of $f_a$.
In Sec.~\ref{sec:axion1}, we review the extra-dimensional one-form axion~\cite{Choi:2003wr}, focusing on the geometric origin of the $2\pi$-periodicity of the 4D QCD axion, its continuous shift symmetry, and how the axion decay constant $f_{a}$ is determined by the warped geometry. Sec.~\ref{sec:axion2}, traces the origin of this shift symmetry to the UV structure, clarifying when it is preserved or violated, and Sec.~\ref{sec:axion3}, discusses a possible 4D holographic dual description of the extra-dimensional axion model in the context of a 5D warped orbifold GUT.
In Sec.~\ref{sec:fa1}, we determine how small $f_{a}$ can be for an axion coupled to a perturbative GUT sector in the AdS$_5$ background, while in Sec.~\ref{sec:fa2}, we show that small-instanton effects remain negligible, even when tree-level contributions make the GUT gauge coupling parametrically large. Sec.~\ref{sec:conclusion} summarizes our results and outlines their implications. Finally, for completeness, in Appendix~\ref{app:c5profile}, we derive the zero mode solution of the 5D axion and present the derivation of the axion decay constant, $f_{a}$ in Appendix~\ref{app:fa}.

In this work, we consider a slice of AdS$_5$ written in conformal coordinates $x_{M} = (x_{\mu}, z)$, with the metric
\ba
ds^{2} &=& g_{MN}\, dx^{M} dx^{N}\cr\cr
&=& \left(\frac{L}{z}\right)^{2} (dx^{2} + dz^{2})
= e^{-2ky} dx^{2} + dy^{2}\,,
\label{eq:metric}
\ea
where $L \equiv k^{-1}$ is the AdS curvature length and $\omega(z) \equiv L/z$ the warp factor. The corresponding vielbein is $e^{A}_{M} = \omega(z)$, satisfying 
$g_{MN} = e_{M}^{A} e_{N}^{B} \eta_{AB} = \omega^{2}(z)\, \eta_{MN}$, 
with the flat metric $\eta_{MN} = \mathrm{diag}(-1, +1, +1, +1, +1)$. In (\ref{eq:metric}), we have also introduced the coordinate $y \in [0, \pi R]$, corresponding to the compactification of the fifth dimension on the orbifold $S^{1}/\mathbb{Z}_{2}$ with radius $R$. The two coordinate systems are related by
\begin{align}
\frac{L}{z} = e^{-k y}\,,
\label{eq:yzrelation}
\end{align}
and will be used interchangeably as convenient for subsequent analysis. The AdS$_5$ line element is invariant under the scaling transformation $x^{\mu} \to \lambda x^{\mu}$ and $z \to \lambda z$, which is equivalent to a constant shift in $y$ and encodes the conformal isometry of the background. 

The orbifold boundaries host a UV brane at $z_{\rm UV} = L$ and an IR brane at $z_{\rm IR} = L e^{\pi k R}$. The 5D fundamental scale $M_{5}$ is related to the 4D (reduced) Planck mass $M_{P} = 2.4 \times 10^{18}\,\mathrm{GeV}$ by~\cite{Randall:1999ee}
\begin{align}
M_{P}^{2} = \frac{M_{5}^{3}}{k}\,\big(1 - e^{-2\pi k R}\big).
\label{eq:M5Mp}
\end{align}

\section{5D GUT in a slice of AdS$_5$}\label{sec:5DGUT}
The underlying idea of grand unification in 4D is that the three gauge interactions of the SM characterized by $SU(3)_{c}\times SU(2)_{L}\times U(1)_{Y}$ arise as low-energy remnants of a single gauge group $G_{\rm GUT}$ that is spontaneously broken at a high scale $M_{\rm GUT}$. In the minimal realization based on $SU(5)$~\cite{Georgi:1974sy}, the SM gauge fields are unified into the adjoint representation ${\bf 24}$, and the SM fermions of each generation fit neatly into ${\bf 10}\oplus\bar{\bf 5}$ representations.

The three Standard Model gauge couplings 
$g_3, g_2, g_1$ originate from a single coupling $g_{\rm GUT}$ at the unification scale $M_{\text{GUT}}$. The one-loop evolution of $\alpha_i=g_i^2/(4\pi)$,
\be
\mu \, \frac{d \alpha^{-1}_i}{d\mu} = \frac{b_i}{2\pi},
\ee
with $b_i = (b_{3},b_{2},b_{1}) = (7,\,19/6,\,-41/10)$ in the SM, shows that $\alpha_1$ and $\alpha_2$ nearly meet around 
$10^{13}\,\mathrm{GeV}$, while $\alpha_3$ fails to unify, rendering the minimal non-supersymmetric 
$SU(5)$ inconsistent with precise gauge coupling unification and proton decay bounds. 
In the MSSM, the beta-function coefficients change to 
$b_i = (3,\,-1,\,-33/5)$, leading to an accurate convergence of all three couplings at 
$M_{\text{GUT}}^{(\mathrm{MSSM})} \simeq 2 \times 10^{16}\,\mathrm{GeV}$ with 
$\alpha_{\text{GUT}}^{-1} \simeq 24$~\cite{Dimopoulos:1981yj,Ibanez:1981yh,Sakai:1981gr}. 
This success of unification remains a central motivation for supersymmetric (SUSY) GUTs together with the prediction for $\sin^{2}\theta_{w}\simeq0.231$, despite ongoing 
constraints from proton decay and threshold effects near the GUT scale.

In 4D SUSY GUTs, the spontaneous breaking of $SU(5)$ is achieved through the vacuum expectation value (VEV) of an adjoint scalar field, $\mathbf{24}_{\Phi}$. However, when the framework is extended to a 5D gauge theory, the unification symmetry can be broken not only by a scalar VEV but also through non-trivial orbifold parity assignments. The corresponding parity eigenvalues at the UV and IR branes determine the boundary conditions of bulk fields, and different eigenvalues for the MSSM gauge multiplets and $X,Y$ gauge multiplets can lead to the breaking of $SU(5)$. While this orbifold breaking mechanism elegantly addresses the doublet--triplet splitting problem, it does not automatically enforce $SU(5)$-symmetric couplings for brane-localized operators. Therefore, achieving gauge coupling unification and successful matching to low energy data generally requires additional assumptions regarding these boundary couplings~\cite{Goldberger:2002cz,Goldberger:2002hb,Goldberger:2002pb}. Alternatively, one may introduce a chiral multiplet $\mathbf{24}_{\Phi}$ localized on the UV brane, whose scalar component acquires a VEV of order $10^{16}\,{\rm GeV}$, thereby inducing the spontaneous breaking $SU(5) \rightarrow SU(3)_{c}\times SU(2)_{L}\times U(1)_{Y}$~\cite{Pomarol:2000hp}.

\begin{figure}[t]
\centering
\hspace*{-5mm}
\includegraphics[width=0.5\textwidth]{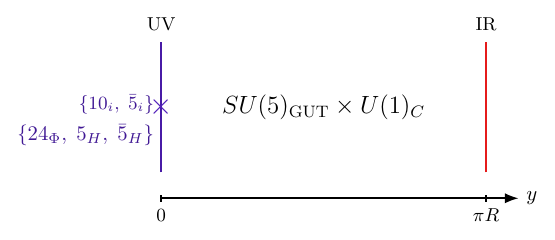}
\caption{Schematic plot showing the field content of the extra-dimensional one-form QCD axion in a 5D warped orbifold GUT. The {\bf{\color{violet}purple}} and {\bf{\color{red}red}} lines denote the {\color{violet}UV} and {\color{red}IR} branes located at $y=0$ and $y=\pi R$ respectively. The bulk contains the $SU(5)_{\rm GUT}$ and axion $U(1)_{C}$ vector supermultiplets, while the MSSM chiral multiplets $\{10_i,\;\bar{5}_i\}$ and Higgs fields $\{24_{\Phi},\;5_H,\;\bar{5}_H\}$ are localized on the UV brane.}
\vspace*{-1.5mm}
\label{fig:RSbraneworld}
\end{figure}
%

In both configurations, the zero modes of the gauge fields constitute an $\mathcal{N}=1$ vector multiplet $V = (A_{\mu}, \lambda)$, whereas their KK excitations assemble into an $\mathcal{N}=2$ vector multiplets $\{V, \Sigma\}$, where the chiral superfield $\Sigma = ((\sigma + iA_5)/\sqrt{2}, \lambda')$ contains the fifth component of the gauge field and its superpartner. For simplicity, we focus on the setup in which the adjoint chiral multiplet $\mathbf{24}_{\Phi}$ is localized on the UV brane ($y=0$), and its VEV triggers the spontaneous breaking of $SU(5)$. The matter sector---comprising the chiral superfields of quarks and leptons $(10_i, \bar{5}_i)$ and the Higgs multiplets $(5_H, \bar{5}_H)$---is also assumed to reside on the UV brane. In the bulk, one finds (1) $\mathcal{N}=1$ vector multiplets corresponding to the zero modes of the MSSM gauge bosons $(V_{321}^{(0)})$ and the $X,Y$ gauge bosons $(V_{X,Y}^{(0)})$, and (2) $\mathcal{N}=2$ vector multiplets describing the KK excitations of these fields, namely $\{V_{321}^{(n\ge1)}, \Sigma_{321}^{(n\ge1)}\}$ and $\{V_{X,Y}^{(n\ge1)}, \Sigma_{X,Y}^{(n\ge1)}\}$. The setup is schematically illustrated in Fig.~\ref{fig:RSbraneworld}. Differing from breaking the GUT group with the orbifold boundary conditions, there is no concern on spoiling the unification due to the dimensionless coupling constants of UV brane-localized operators as the UV brane respects $SU(5)$.\footnote{While the UV-localized renormalizable gauge kinetic term $\propto{\rm Tr}[G_{\mu\nu}G^{\mu\nu}]\delta(y-y_{\rm UV})$ respects $SU(5)_{\rm GUT}$ and thus does not spoil the gauge unification, higher-dimensional, UV-localized operators, such as $\delta(y-y_{\rm UV}){\rm Tr}[{\bf 24}_{\Phi}G_{\mu\nu}G^{\mu\nu}]/M_{P}$, can induce minor threshold corrections to the differences in four-dimensional gauge couplings.}

In a 5D gauge theory, the running behavior of the gauge coupling critically depends on the geometry of the extra dimension. In the case of a flat extra dimension, the gauge field admits a KK 
tower of massive excitations whose threshold effects alter the RG evolution of the 
4D effective gauge coupling. These KK modes contribute 
at each mass threshold, leading to an effective power-law running of the gauge coupling above the compactification scale $R^{-1}$~\cite{Dienes:1998vh,Dienes:1998vg}. Based on this observation, often
maintaining perturbativity of $g_{\rm GUT}$ at $\mu=M_{\rm GUT}$ leads to the requirement that the compactification scale, 
or first KK mass $m_{\rm KK}\!\simeq\! R^{-1}$, lie sufficiently close to the unification scale, 
setting a lower bound on $m_{\rm KK}\gtrsim M_{\rm GUT}$~\cite{PaccettiCorreia:2002gf,Benabou:2025kgx}.

In contrast, when the extra dimension is warped, the effective 4D gauge coupling exhibits only logarithmic running. For $q\lesssim m_{\rm KK}$, only the zero mode remains dynamical at low energies and the zero-mode two-point correlator serves as the appropriate observable for investigating the running of the gauge coupling. As shown  by Pomarol~\cite{Pomarol:2000hp}, the contribution to the massless-mode gauge-boson propagator from KK states of bulk charged fields depends only 
logarithmically on the ultraviolet regulator $M_{\rm PV}$, provided $M_{\rm PV} \lesssim k$. 
This behavior can be traced to the near-degeneracy 
between the KK excitation spectra of the charged bulk fields and their 
Pauli-Villars (PV) regulators: level-by-level cancellation nullifies the naively anticipated KK-induced 
power-law sensitivity to the UV regulator.

For $q\gtrsim m_{\rm KK}$, the zero-mode two-point correlator can be dominated by higher-dimension terms, signaling the breakdown of the EFT~\cite{Goldberger:2002cz,Goldberger:2002hb}. As a result, the zero-mode two-point correlator no longer determines the coefficient of the local gauge-field kinetic term and cannot be used to extract the running gauge coupling (equivalently, in the dual 4D picture, the elementary gauge field mixes strongly with the composite resonances). The relevant observable above the IR scale is instead the Planck-brane gauge correlator $G_p(k^{-1},k^{-1})$ where $G_p(z,z')$ is the Fourier transform of the 5D gauge field propagator and $p=\sqrt{\eta_{\mu\nu}p^{\mu}p^{\nu}}$ with $p^{\mu}$ the 4D four-momentum. This correlator describes the elementary (source) gauge field that weakly gauges the conserved current of the dual CFT, and remains well-defined even when KK modes are kinematically accessible.
At tree level, its the bulk contribution already exhibits logarithmic running of the gauge coupling
\be
\frac{8\pi^{2}}{g_{i}^{2}(
q)}=\frac{8\pi^{2}}{kg_{5}^{2}}\ln\left(\frac{k}{q}\right)+...\,,
\label{eq:treecft}
\ee
where $g_{i}$ is the effective 4D gauge coupling and $g_{5}$ is 5D gauge coupling with the mass dimension $[g_{5}^{-2}]=1$. When applying the AdS/CFT correspondence~\cite{Maldacena:1997re,Gubser:1998bc,Witten:1998qj} to RS1, the coefficient \(b_{\rm CFT}\equiv 8\pi^{2}/(k g_{5}^{2})\) can be understood holographically as the CFT contribution to the gauge coupling running~\cite{Arkani-Hamed:2000ijo}. The dual 4D description of a 5D bulk gauge theory is a strongly-coupled CFT with a large number of CFT ``colors", $N_{\rm CFT}$ and a global symmetry.
The 4D dual theory is described by the Lagrangian
\be
\mathcal{L}_{\rm 4D}=\mathcal{L}_{\rm CFT}-\frac{1}{4g^{2}}F_{\mu\nu}F^{\mu\nu}+A_{\mu}J^{\mu}_{\rm CFT}+\text{h.c.}\,,
\label{eq:L4D}
\ee
where $\mathcal{L}_{\rm CFT}$ describes the (unspecified) CFT that is spontaneously broken at the IR scale $m_{\rm IR}\equiv z_{\rm IR}^{-1}=ke^{-\pi kR}$, $J^{\mu}_{\rm CFT}$ is the global symmetry CFT current, and $A_{\mu}$ is the 4D gauge field that weakly gauges the global symmetry corresponding to the bulk gauge symmetry. Remarkably, $b_{\rm CFT}$ has an “IR-free" sign. It is a universal effect and thus equally contributes to the running of all 4D effective gauge couplings $g_{i}$. 

On the other hand, since the bulk-to-brane 
propagator satisfies 
$G_p(z,k^{-1}) \propto e^{-pz}$ for $z$ near $z_{\rm IR}$ and $pz\gg1$, radiative corrections 
to the Planck-brane correlator are dominated by the region near $z = k^{-1}$~\cite{Goldberger:2002cz,Goldberger:2002hb}. Because the 
zero-mode wave function is localized towards $z=k^{-1}$, whereas the KK-excited states with 
$m_{\mathrm{KK}} \ll k$ are localized toward the IR brane (large $z$), the resulting loop integral is governed by UV-localized zero-modes, 
yielding the familiar logarithmic contribution to the running of the 4D effective gauge coupling with 
coefficient $b_i$. 

Combining the above effects leads to the following {\it logarithmic} running of the 4D effective gauge couplings 
\ba
\frac{8\pi^{2}}{g_{4,i}^{2}(q)}
=\frac{8\pi^{2}}{g_{4,i}^{2}(m_{Z})}-b_{i}\ln\left(\frac{m_{Z}}{q}\right)+b_{\rm CFT}\ln\left(\frac{m_{\rm KK}}{q}\right)\,,
\label{eq:g4higher}
\ea
where $m_{Z}=91.2~{\rm GeV}$ is the $Z$-boson mass, $\alpha^{-1}_{1}(m_{Z})=59.1$, $\alpha^{-1}_{2}(m_{Z})=29.6$, $\alpha^{-1}_{3}(m_{Z})=8.47$ and the CFT contribution parametrized by $b_{\rm CFT}$ is absent for $q<m_{\rm KK}$.  Because of the term $A_{\mu}J^{\mu}_{\rm CFT}$ in (\ref{eq:L4D}), \(b_{\rm CFT}\) effectively counts the number of CFT degrees of freedom~\cite{Arkani-Hamed:2000ijo,Randall:2001gb,Pomarol:2000hp,Contino:2002kc}. This implies $b_{\rm CFT}\simeq N_{\rm CFT}$ and $c\simeq N_{\rm CFT}^{2}$ where $c$ is the central charge  of the dual 4D CFT.

\begin{figure}[t]
\centering
\hspace*{-5mm}
\includegraphics[width=0.53\textwidth]{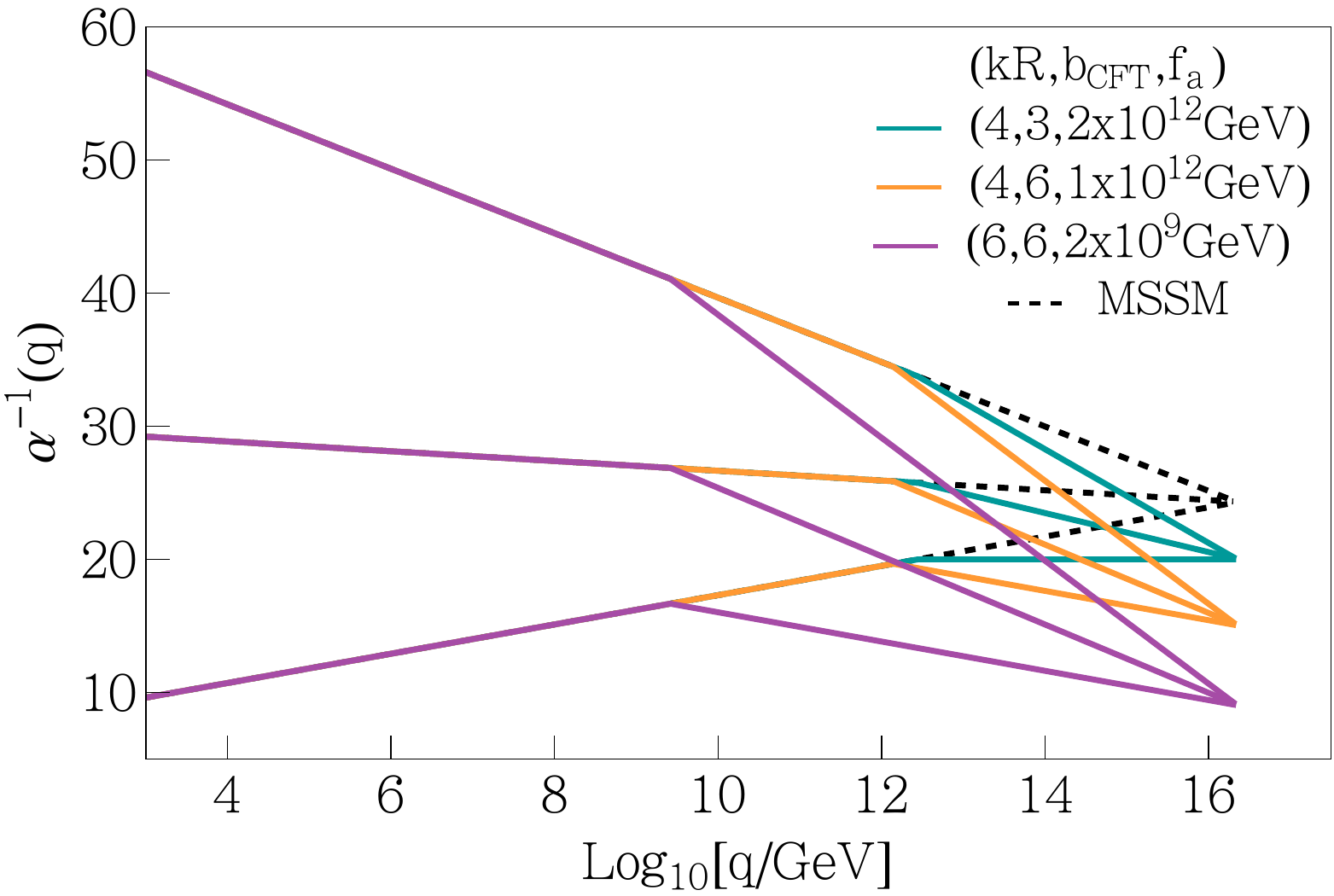}
\caption{The Standard Model couplings $\alpha^{-1}_{i}=(4\pi)/g_{i}^{2} \,\,(i=1,2,3)$ as a function of $q$ for several values of $f_a$ and $b_{\rm CFT}$. The {\bf{\color{violet} purple}}, {\bf{\color{orange}orange}} and {\bf{\color{teal}green}} solid lines correspond to $\alpha_i^{-1}(q)$ for $(kR,b_{\rm CFT},f_{a})= {\color{violet}(6,6,5\times10^{9}~{\rm GeV})},{\color{orange}(4,6,2\times10^{12}~{\rm GeV})}$ and ${\color{teal}(4,3,4\times10^{12}~{\rm GeV})}$ respectively, while the {\bf black} dashed lines show the MSSM running.
}
\vspace*{-1.5mm}
\label{fig:alphalnt}
\end{figure}

Additionally, $c$ can be determined holographically from the two-point correlator of the stress-energy tensor, \(\langle T_{\mu\nu}(x) T_{\rho\sigma}(0)\rangle\), which is generated by the exchange of the 5D graviton in the AdS bulk~\cite{Henningson:1998gx,Freedman:1999gp,Arutyunov:1999nw}. In the AdS/CFT correspondence, the normalization of this correlator is governed by the kinetic term of the 5D graviton, and hence by the 5D Planck scale $M_{5}$. More precisely, one finds $c\sim M_{5}^{3}/k^{3}$. Using \(M_{P}^{2}\simeq M_{5}^{3}/k\)~\cite{Randall:1999ee} and $c\simeq N_{\rm CFT}^{2}$, then one finds $N_{\rm CFT}\sim \sqrt{c}\sim M_{P}/k$. Thus, for typical warped scenarios with \(k\sim 0.1\,M_{P}\), this yields \(b_{\rm CFT}=\mathcal{O}(10)\), consistent with the interpretation of \(b_{\rm CFT}\) as the effective large-\(N\) coefficient of the CFT contribution to the gauge coupling running.

In Fig.~\ref{fig:alphalnt}, we present the RG evolution of the inverse gauge couplings, $\alpha_i^{-1} = 4\pi / g_i^2$, as a function of $q$. The black dashed lines represent the evolution of the MSSM gauge couplings, which unify at $M_{\rm GUT} \simeq 2 \times 10^{16}\,{\rm GeV}$. The purple, orange, and green lines correspond respectively to parameter choices $(kR, b_{\rm CFT},f_{a}) = (6,6,5\times10^{9}~{\rm GeV})$, $(4,6,2\times10^{9}~{\rm GeV})$, and $(4,3,4\times10^{12}~{\rm GeV})$ where the dimensionless quantity $kR$ determines the KK mass scale, $m_{\rm KK} \sim \pi k e^{-\pi kR}$, while $M_P / k \simeq N_{\rm CFT} \simeq b_{\rm CFT}$ parametrizes the tree-level contribution (interpreted to be a CFT contribution). The gauge coupling unification at $M_{\rm GUT} \simeq 2 \times 10^{16}\,{\rm GeV}$ persists even when an additional contribution $b_{\rm CFT} = \mathcal{O}(10)$ to the running of $g_i$ turns on for $q \gtrsim m_{\rm KK}$. This is because the tree-level (CFT) contribution is universal, shifting all $g_i$ equally and leaving the relative differences between any pair of gauge couplings unchanged from their standard 4D GUT values.


\section{extra dimensional (one-form) Axion}\label{sec:axion}
\subsection{4D axion from $C_{5}$}
\label{sec:axion1}

In the extra-dimensional one-form axion framework~\cite{Choi:2003wr}, the fifth dimension is compactified on an $S^{1}/\mathbb{Z}_{2}$ orbifold. The bulk not only contains the MSSM gauge supermultiplets but also an additional 5D $U(1)_{C}$ gauge multiplet, whose gauge symmetry is interpreted as the Peccei-Quinn symmetry in the dual 4D picture. The $U(1)_{C}$ gauge multiplet decomposes into a 4D  $\mathcal{N}=1$ vector multiplet $V_{c} = (C_{\mu}, \lambda_{c})$ and a chiral multiplet 
$\Sigma_{c} = ((\sigma_{c} + i C_{5}), \lambda_{c}')$. 
The 5D gauge field $C_{M}$ obeys the periodicity condition $C_{M}(y) = C_{M}(y + 2\pi R)$ on the covering circle.

We impose Dirichlet boundary conditions on $V_{c}$ and Neumann boundary conditions 
on $\Sigma_{c}$ at both branes, ensuring that only the chiral multiplet $\Sigma_c$ has a zero 
mode. Consequently, the 4D low-energy spectrum contains an axion as well as its scalar partner (the saxion), and the corresponding fermionic axino. After the addition of suitable supersymmetry-breaking terms on the UV brane, the saxion and axino acquire masses of order the gravitino mass $m_{3/2}$, placing them well above the axion in the spectrum. As shown in Appendix~\ref{app:c5profile}, from the boundary conditions and the gauge-fixing term that removes the mixing between $C_{\mu}$ and $C_{5}$, the zero mode solution of $C_{5}(x,y)$ is independent of the extra-dimensional coordinate $y$. However, the kinetic term of $C_5$ contains warp factors so that with respect to the 5D flat metric the zero-mode profile is proportional to $e^{k y}$, implying that the mode is localized toward the IR brane~\cite{Flacke:2006ad} (see Appendix~\ref{app:fa}).

After dimensional reduction, the zero mode of $C_{5}$ gives rise to a 4D scalar $\theta_{a}(x)$ identified with the Wilson line (holonomy) of the 5D gauge field around the compact dimension
\be
\theta_{a}(x)=\frac{a(x)}{f_{a}}=\oint dy\,C_{5}(x,y)=2\pi RC_{5}(x)\,,
\label{eq:thetaax}
\ee
where $a(x)$ is the 4D axion field and $f_a$ the axion decay constant. Under the $U(1)_{C}$ symmetry, the 5D gauge field transforms as $C_{M} \rightarrow C_{M} + \partial_{M}\xi$ with the gauge parameter $\xi$ obeying $\xi(y)=\xi(y+2\pi R)\,\,\,{\rm mod}\,\,\,2\pi$ and $\xi(-y)=-\xi(y)\,\,\,{\rm mod}\,\,\,2\pi$. In particular, the 5th component transforms as $C_{5} \rightarrow C_{5} + \partial_{5}\xi$. For small gauge transformations with $\xi_{S}(y+2\pi R) = \xi_{S}(y)$, the Wilson line $\theta_a$ is invariant. By contrast, large gauge transformations obeying  $\xi_{L}(y+2\pi R) = \xi_{L}(y) + 2\pi n$ with $n\neq0$ and $n\in\mathbb{Z}$, shift $\theta_{a}(x)$ by an integer multiple of $2\pi$. For example, consider $\xi_{L}(y)=ny/R$ where it can be immediately seen that 
\be
\delta_{\xi_{L}}\theta_{a}(x)=\oint dy\,\partial_{5}\xi_{L}=2\pi n,\quad n\in\mathbb{Z}\,,
\ee
demonstrating that $\theta_{a}(x)$ is a compact angular variable with periodicity $2\pi$.

The $U(1)_{C}$ gauge symmetry can be used to constrain the possible local bulk operators involving $C_M$. In the absence of matter fields charged under $U(1)_{C}$, the only allowed gauge-invariant terms are the kinetic term for $C_{M}$ and its Chern-Simons coupling to other gauge sectors (where $\xi|_{\partial {\rm AdS}_{5}}=0$).
Accordingly, when the extra-dimensional axion model is embedded in a 5D orbifold GUT framework, 
$C_{M}$ appears in the 5D action as\footnote{ In odd spacetime dimensions, gauge theories admit local Chern–Simons forms, $\omega_{2n+1}(A)$ $(n\in\mathbb{Z})$, defined by ${\rm d}\omega_{2n+1}(A)={\rm Tr}[F^{n+1}]$. While $\omega_{2n+1}$ is not strictly gauge invariant, its gauge variation is an exact form, so the associated action is gauge invariant up to boundary terms. This mechanism is special to odd spacetime dimensions and has no analogue in even dimensions. Consequently, 5D effective theories can contain intrinsic Chern–Simons–type couplings, such as $\propto C\wedge{\rm Tr}[G\wedge G]$, without requiring fermionic matter or invoking quantum anomalies.}
\ba
S_{\rm 5D}
&=&\int d^4 x\int_{y_{\rm UV}}^{y_{\rm IR}}dy\sqrt{-g}\,\,\left\{\frac{1}{4g_{5C}^{2}}C_{MN}C^{MN}\right.\nonumber\\
&&\left.+\frac{1}{\sqrt{-g}}\frac{c_{5}}{16\pi^{2}}\epsilon^{MNRST} C_{M}{\rm Tr}[G_{NR}G_{ST}]+...\right\},\,\nonumber\\
\label{eq:S5D}
\ea
where $C_{MN}$ and $G_{MN}$ denote the field strength of $C_{M}$ and the GUT gauge field $A_{M}$, respectively, 
and $c_{5}\in\mathbb{Z}$ is an anomaly coefficient. The ellipses denote the kinetic terms for the fields contained in the $SU(5)$ 
gauge supermultiplets and $U(1)_{C}$ gauge supermultiplet other than $C_{M}$ as well as terms involving the saxion $\propto\sigma_{c}{\rm Tr}[G^{MN}G_{MN}]$ and $\propto\sigma_{c}\partial_{M}\sigma_{c}\partial^{M}\sigma_{c}$~\cite{Arkani-Hamed:2001vvu}.

The 4D axion shift symmetry can be understood as descending from the electric one-form symmetry $U(1)_{e}^{[1]}$ of the 5D $U(1)_{C}$ gauge theory~\cite{Reece:2023czb}. Under the $U(1)_{e}^{[1]}$ symmetry, generated by the two-form current $j_{e}^{[2]} = \frac{{\rm d}C}{g_{5}^{2}}$, the gauge field shifts by a closed one-form connection $\Lambda_{e}^{[1]}$ satisfying $\oint \Lambda_{e}^{[1]} \in S^{1}$. In particular, the fifth component transforms as $C_{5} \rightarrow C_{5} + \Lambda_{e,5}$, where $\Lambda_{e,5}$ denotes the fifth component of  $\Lambda_{e}^{[1]}$. Thus, upon dimensional reduction, the $S^{1}$-component of this transformation induces a shift in the 4D axion field,
\be
\theta_{a}(x) \;\rightarrow\; \theta_{a}(x) + \oint dy\, \Lambda_{e,5} 
\;=\; \theta_{a}(x) + \alpha\,,
\label{eq:4Dshift}
\ee
with $\alpha \in [0, 2\pi)$. Thus, the continuous shift of the axion is simply the 4D remnant of the $U(1)_{e}^{[1]}$ symmetry.

The familiar 4D ABJ anomaly associated with the shift symmetry in (\ref{eq:4Dshift}) is actually traced back to the 5D equation of motion for $C_{M}$, which determines the divergence of the electric symmetry current, ${\rm d}\star j_{e}^{[2]}\propto c_{5}\,{\rm Tr}\!\left[G \wedge G\right]$. Integrating the zero modes on both sides of this equation over $y$ yields the 4D ABJ anomaly of the 4D axion shift symmetry with respect to the GUT gauge group. Thus, the (anomalous) nonconservation of the 4D PQ current is simply the dimensional reduction of the Chern-Simons induced violation of the 5D electric one-form symmetry.

Finally, the axion decay constant $f_a$ can be obtained by substituting the zero mode solution into (\ref{eq:S5D}) and canonically normalizing the 4D axion kinetic terms (see Appendix~\ref{app:fa}) to give
\ba
f_{a}
=\sqrt{\frac{k}{4g_{5C}^{2}}\,\frac{1}{e^{2\pi kR}-1}}\simeq \frac{1}{2}  N_{\rm CFT}^{\frac{1}{3}}m_{\rm IR},
\label{eq:fa}
\ea
where $g_{5C}^{-2} \simeq M_{5}$ and $M_{P}/k\simeq N_{\rm CFT}$ were used in the second relation. This result shows that the axion decay constant is naturally tied to the IR scale and the effective number of CFT degrees of freedom.

\subsection{Extra-dimensional axion quality}
\label{sec:axion2}

Because the QCD axion arises from $C_{5}$ after 
dimensional reduction, any non-derivative interaction of $C_{M}$ induces a contribution to the 
4D axion potential. When no charged matter propagates in the bulk, the Chern-Simons term in (\ref{eq:S5D}) is the only allowed non-derivative interaction for $C_M$ (and hence $C_5$). This constraint follows directly from the interplay of $U(1)_C$ gauge invariance and the requirement of 5D locality~\cite{Choi:2003wr}. Consequently, in a bulk devoid of charged fields, the divergence of the electric two-form current, ${\rm d}\star j_{e}^{[2]}$, receives contributions solely from the mixed anomaly between $U(1)^{[1]}_{e}$ and the GUT gauge sector, i.e. ABJ anomaly.  A systematic classification of possible sources for ${\rm d}\star j_{e}^{[2]}\neq0$ can be found in~\cite{Craig:2024dnl}.

The absence of any additional contribution to ${\rm d}\star j_{e}^{[2]}$ beyond the ABJ anomaly implies that the 4D axion mass is generated exclusively by the nonperturbative $SU(3)_{c}$ dynamics. Owing to the boundary conditions imposed on $C_{\mu}$ and $C_{5}$, the $S^{1}$ component of the bulk one-form symmetry $U(1)_{e}^{[1]}$ descends to a zero-form shift symmetry $U(1)_{e}^{[0]}$ acting on the axion (as discussed in Sec.~\ref{sec:axion1}), while the $\mathbb{R}^{1,3}$ component of $U(1)_{e}^{[1]}$ is explicitly broken at the boundaries~\cite{Damia:2022bcd}. Consequently, upon dimensional reduction, ${\rm d}\star j_{e}^{[2]}$ reduces to ${\rm d}\star{\rm d}\theta_{a}$. As long as $U(1)_{e}^{[1]}$ is violated only by the ABJ anomaly, and its boundary projection $({\rm i.e.}~U(1)_{e}^{[0]})$ robustly forbids any local operators of the form $(C_5)^p$ with integer $p \ge 2$ at the orbifold fixed points, the shift symmetry $U(1)_{e}^{[0]}$ is broken solely by the 5D Chern-Simons coupling between $C_{M}$ and the GUT gauge fields and there is no quality problem for the extra-dimensional axion.\footnote{One might worry that small instanton effects could jeopardize axion quality by introducing misaligned contributions to its potential. However, as will be shown in Sec.~\ref{sec:fa2}, these effects are in fact harmless and do not spoil the axion solution to the strong CP problem.} 

By contrast, once a charged bulk field is introduced, its very presence explicitly breaks $U(1)_{e}^{[1]}$ and immediately revives the axion-quality problem: charged matter inevitably sources an additional
contribution to ${\rm d}\!\star j_{e}^{[2]}$ besides the ABJ anomaly. Specifically, let us consider a bulk complex scalar $\Phi$ of mass $m_{\Phi}$ and charge $q_{\Phi}$ under $U(1)_{C}$, with 5D action
\be
S_{\rm 5D}^{\Phi}
=\int d^4 x\int_{y_{\rm UV}}^{y_{\rm IR}}dz\sqrt{-g}\,\,\left(|D_{M}\Phi|^{2}+m_{\Phi}^{2}|\Phi|^{2}\right)\,.
\ee
Since $U(1)_{C}$ gauge invariance requires $C_{M}$ to appear through the covariant derivative $D_{M}\Phi=(\partial_{M}-iq_{\Phi}C_{M})\Phi$, the equation of motion for $C_{M}$ necessarily involves $\Phi$. Consequently, the conservation equation for $j^{[2]}_{e}$ is violated, i.e. ${\rm d}\star j_{e}^{[2]} \neq 0$. At the same time, a Wilson line can now end on $\Phi$, leading to the gauge-invariant composite operator
\be
\Phi(x'_{M})\exp\!\left(iq_{\Phi}\!\int_{\gamma} C \right)\,,
\ee
for any path $\gamma$ terminating at $x'_{M}$. Because the would-be symmetry defect of $U(1)_{e}^{[1]}$ no longer acts consistently on such composites, it fails to produce a well-defined action, precisely reflecting the explicit breaking of the electric one-form symmetry.

Indeed, once the field $\Phi$ is introduced, it can generate $\theta_{a}$-dependent 
contributions to the vacuum energy---through both classical (tree-level) and quantum effects---thereby inducing a nontrivial additional 4D axion potential, $\delta V(\theta_{a})$. To illustrate the possible tree-level contribution~\cite{Petrossian-Byrne:2025mto,Deshpande:2019kjl}, let us consider brane-localized operators $\mathcal{O}_{\rm UV}$ and $\mathcal{O}_{\rm IR}$ that transform as $\mathcal{O}_{\rm UV}\rightarrow\mathcal{O}_{\rm UV}e^{iq_{\Phi}\xi(y_{\rm UV})}$ and $\mathcal{O}_{\rm IR}\rightarrow\mathcal{O}_{\rm IR}e^{-iq_{\Phi}\xi(y_{\rm IR})}$ under the projection of the $U(1)_{C}$ gauge symmetry onto each brane. Gauge invariance then permits $\Phi$ to couple to these operators through the brane-localized interactions
\ba
\Delta S_{\rm 5D}^{\Phi}&=&\int d^4 x\int_{y_{\rm UV}}^{y_{\rm IR}}dy\sqrt{-g}\,\,\{\kappa_{\rm IR}
\mathcal{O}_{\rm IR}\Phi\delta(y-y_{\rm IR})\cr\cr
&&+\kappa_{\rm UV}\mathcal{O}_{\rm UV}\Phi^{\dagger}\delta(y-y_{\rm UV})\}+{\rm h.c.}\,,
\label{eq:DeltaS5DPhi}
\ea
where the dimensionless couplings $\kappa_{\rm UV,IR}$ are in general complex. Since (\ref{eq:DeltaS5DPhi}) acts as a source for the $\Phi$ equation of motion, integrating out $\Phi$ at tree-level generates the nonlocal term $\delta V(\theta_{a})\propto|\kappa_{\rm IR}||\kappa_{\rm UV}|\mathcal{O}_{\rm IR}\mathcal{O}_{\rm UV}e^{-\nu\pi k R}{\rm exp}\left(iq_{\Phi}\int_{y_{\rm UV}}^{y_{\rm IR}}{\rm d}y\,C_{5}+i\delta_{\kappa}\right)+{\rm h.c.}$ with $\nu\equiv\sqrt{4+\frac{m_{\Phi}^{2}}{k^{2}}}$ and $\delta_{\kappa}$ the relative phase between $\kappa_{\rm IR}$ and $\kappa_{\rm UV}$.\footnote{Here the physical axion corresponds to the overall phase of the gauge-invariant nonlocal operator $\mathcal{O}_{\rm UV}{\rm exp}\left(iq_{\Phi}\int_{y_{\rm UV}}^{y_{\rm IR}}{\rm d}y\,C_{5}\right)\mathcal{O}_{\rm IR}$.} Thus, whenever a charged bulk field couples in a gauge-invariant manner to brane-localized operators, it generically induces a tree-level contribution $\delta V(\theta_{a})$, which is not removed by supersymmetry. Nevertheless, such effects can be rendered parametrically small in the limit $\nu k\gg R^{-1}$, where the exponential suppression becomes significant.

In the absence of brane-localized interactions such as (\ref{eq:DeltaS5DPhi}), one can indeed evade any tree-level contribution to $\delta V(\theta_{a})$. Nevertheless, the story does not end there: the minimal coupling of $\Phi$ to $C_{M}$ inevitably leads to a radiatively generated effective (Casimir) potential for the line operator, $e^{iq_{\Phi}\int C_{5}}$~\cite{Hosotani:1983xw,Cheng:2002iz,Contino:2003ve,Arkani-Hamed:2007ryu,Reece:2023czb}. The one-loop effective potential for the Wilson line arises from summing the vacuum energies (equivalently, functional determinants) of the entire KK tower of $\Phi$, i.e.
\be
\delta V(\theta_{a})\simeq\sum_{n}\int\frac{{\rm d^4} p}{(2\pi)^{4}}\ln[p^{2}+m_{{\rm KK},n}^{2}(\theta_{a})]\,,
\label{eq:1loopnonSUSY}
\ee
where $n$ labels the KK states and $m_{{\rm KK},n}(\theta_{a})$ is a $\theta_{a}$-dependent mass. In a flat extra dimension, this contribution scales as $e^{-2\pi m_{\Phi}R}$, leading to exponentially strong suppression for $m_{\Phi}\gg R^{-1}$~\cite{Reece:2023czb}. An analogous computation in a warped extra-dimensional background proceeds using the $\theta_{a}$-dependent KK spectrum appropriate to the warped geometry. Interestingly, this setting produces (i) a larger worldline instanton action and (ii) an additional suppression factor $m_{\rm KK}^{4}\propto e^{-4\pi k R}$, reflecting the low-lying KK masses in the warped case~\cite{ChoiTalk2025,Choi:work}. In our setup, even when a charged matter field propagates in the bulk, both the tree-level contribution and the loop-induced effects can be made parametrically 
small as long as the hierarchy $\nu k \gg R^{-1}$ is maintained. In this regime, the warped background strongly suppresses the radiatively-induced $\delta V(\theta_{a})$, thereby significantly improving axion quality relative to the flat space case.

Furthermore, beyond the suppression due to the warped geometry, supersymmetry adds another layer of protection against quantum corrections. Even though SUSY breaking removes the exact boson-fermion cancellation in the vacuum energy, sizable partial cancellations remain, greatly reducing the radiative corrections relative to the non-supersymmetric case. At one loop, the additional contribution to the 4D axion potential takes the schematic form
\ba
\delta V(\theta_{a})&\simeq&\sum_{n}\int\frac{{\rm d^4}p}{(2\pi)^{4}}\left\{ \ln[p^{2}+m_{{\rm KK},n}^{(b)\,2}(\theta_{a})]\right.\cr\cr
&&\qquad\qquad-\left.\ln[p^{2}+m_{{\rm KK},n}^{(f)\,2}(\theta_{a})]\right\}\,,
\label{eq:1loopSUSY}
\ea
where $m_{{\rm KK},n}^{(b,f)}$ are the $\theta_{a}$-dependent KK masses of $\Phi$ and its (fermionic) superpartner. In particular, whenever $m_{{\rm KK},n}^{(b)}\approx m_{{\rm KK},n}^{(f)}$, the $n$-th KK level negligibly contributes to $\delta V(\theta_{a})$ due to the near-perfect boson–fermion cancellation. In our setup, SUSY breaking is localized on the UV brane, so the soft mass of the $n$-th KK mode scales as $m_{\rm soft}^{2}\propto m_{3/2}^{2}|f_{\Phi}^{(n)}(y_{\rm UV})|^{2}$, where $f_{\Phi}^{(n)}(y_{\rm UV})$ is the n-th KK profile of $\Phi$ evaluated at the UV brane. Because the low-lying KK modes are strongly suppressed at $y=y_{\rm UV}$, their SUSY-breaking mass splittings are tiny, and hence their contributions to (\ref{eq:1loopSUSY}) cancel almost exactly. Only the higher KK modes—whose profiles grow toward the UV brane—acquire sizable SUSY-breaking masses and therefore dominate the residual effect. Consequently, the partial cancellations ensure that the radiatively generated $\delta V(\theta_{a})$ in (\ref{eq:1loopSUSY}) is much smaller compared to (\ref{eq:1loopnonSUSY}) for the non-supersymmetric scenario. This combined effect of warping and supersymmetry substantially improves the axion quality of the extra-dimensional axion when embedded in 5D warped orbifold supersymmetric GUTs.

\subsection{4D dual description}
\label{sec:axion3}

Using the AdS/CFT dictionary, we next discuss the possible 4D dual description. The 5D bulk gauge field $C_M$ is dual to a conserved current $J_\mu$ of the 4D CFT.
The boundary conditions imposed on $C_{M}$ determine whether the zero modes of $C_{5}$ and $C_{\mu}$ are present in the low-energy spectrum. The presence of a normalizable zero mode for $C_5$ but the absence of a zero mode for $C_\mu$ implies that the low-energy theory contains a pNGB associated with the spontaneous breaking of the global $U(1)_C$ symmetry of the CFT, but no dynamical photon gauging this symmetry. Since there is no elementary UV gauge field $C_\mu$ acting as a source for the CFT current $J_\mu$, this pNGB is entirely generated by the strong CFT dynamics and is therefore fully composite, with no elementary admixture.

In addition to the global $U(1)_{C}$, the symmetry structure consists of the strongly interacting $SU(N_{\rm CFT})$, and the global $SU(5)_{\rm GUT}$, the latter being weakly gauged through a source field dual to the bulk GUT gauge field $A_{\mu}(y=0)$ on the UV brane. These 4D global symmetries mirror the bulk gauge symmetries of the 5D setup. Notably, this arrangement echoes the original composite axion model 
\cite{Kim:1984pt,Choi:1985cb}, with $SU(N_{\rm CFT})$ naturally playing the role of a confining 
``axicolor'' group. 
 
However, the original 4D composite–axion model—where the axion emerges from a fermion condensate that spontaneously breaks the $U(1)$ PQ symmetry—does not constitute an adequate holographic dual. The reason is that its leading PQ-violating operator can already  appear at dimension 3 (corresponding to a fermion bilinear), 
leaving the axion quality highly vulnerable to UV violation. In fact, this scenario can be modeled in 5D by introducing an additional bulk PQ-charged complex scalar field, dual to the fermion condensate with dimension $\Delta$~\cite{Cox:2019rro}. Not unexpectedly, in this case, the axion quality can be solved by requiring $\Delta\gtrsim 10$. Explicit 4D composite-axion models which realize such high-dimension fermion condensates with naturally massless fermion constituents, have been constructed in Ref.~\cite{Randall:1992ut, Redi:2016esr, Lillard:2018fdt, Gavela:2018paw, Vecchi:2021shj,Contino:2021ayn, Gherghetta:2025fip}.

By contrast, in our 5D construction, where no charged matter is assumed and thus $U(1)_{e}^{[1]}$ is an exact symmetry when the Chern-Simons coupling is absent, the extra dimensional one-form axion is automatically of exceptionally high quality. One natural alternative to match this level of protection in a 4D dual is to identify the axion with the phase of a baryonic operator. Unlike meson-type operators, whose scaling dimensions do not depend on $N_{\rm CFT}$, a baryon operator contains a number of elementary constituents determined by the color and flavor structure of the underlying gauge theory. Consequently, the scaling dimension of a baryonic operator is not fixed, in sharp contrast to that of a meson; it can become parametrically large when the number of colors is taken large. When supersymmetrized, this feature opens the door to a baryonic composite-axion 
framework with naturally high axion quality, since the leading PQ-violating operators are 
pushed to high dimension and thus become extremely suppressed~\cite{Agrawal:2025mke}.

To obtain baryonic operators in the dual 4D CFT, we require a holographic dual that contains degrees of freedom transforming in the fundamental representation of $SU(N_{\rm CFT})$. In a single $SU(N_{\rm CFT})$ gauge theory with only adjoint matter (as in ${\cal N} = 4$ SYM) one can only form trace-type operators, since no determinant-like gauge invariant built purely from adjoints is possible. Although fundamental flavors can be introduced by adding D7-branes in the bulk, their backreaction typically leads to substantial deviation from the AdS$_5$ geometry. Instead, we consider Klebanov-Witten~\cite{Klebanov:1998hh} (or Klebanov-Strassler~\cite{Klebanov:2000hb}) type solutions, whose dual 4D description is an $SU(N_{\rm CFT})_1\times SU(N_{\rm CFT})_2$ gauge theory with bifundamental fields $A_i, B_i$ $(i=1,2)$. These bifundamentals allow the construction of gauge-invariant “determinant” operators by antisymmetrizing over both sets of color indices with epsilon tensors, using $N_{\rm CFT}$ bifundamental insertions\footnote{In SQCD with a single $SU(N_{\rm CFT})$, baryons are antisymmetric in flavor and exist only for $N_f \geq N_{\rm CFT}$, whereas in an $SU(N_{\rm CFT})\times SU(N_{\rm CFT})$ theory with bifundamentals the two color epsilons cancel the sign, so baryons are symmetric in flavor and do not require $N_f \sim N_{\rm CFT}$.}. This realizes a baryonic-type $U(1)_{\rm PQ}$ symmetry with no light charged operators, and the first charged operator appearing at parametrically large dimension $N_{\rm CFT}$. On the gravity side, the Klebanov-Witten background is AdS$_5\times Y_5$, and these baryonic operators are dual to wrapped brane states (e.g. D3 branes wrapping nontrivial cycles of $Y_5$). In the resulting 5D RS effective theory, such wrapped brane states reside at or above the UV cutoff.

The spontaneous breaking of the baryonic-type $U(1)_{\rm PQ}$ in the 4D dual is realized by a condensate of the high-dimension baryonic operator, i.e. by moving onto the baryonic branch of the strongly coupled theory. The phase of this condensate is identified with the composite axion, while the associated radial excitation is a heavy mode ($\propto N_{\rm CFT}$) with mass at or above the cutoff of the RS effective theory. The IR-scale bulk U(1) KK states are mapped in the dual language to a discrete tower of meson-like resonances (i.e. baryon number neutral) in an appropriate confining or baryonic-branch deformation of the CFT (analogous to Klebanov-Strassler-like behavior). To embed $SU(5)_{\rm GUT}$, one can introduce a small number $N_f$ of flavor multiplets into the 4D CFT (e.g. via D7-branes in the UV completion~\cite{Gherghetta:2006yq}), enlarging the global symmetry to include a flavor factor $U(N_f ) \supset SU(5)_{\rm GUT}$, which is then weakly gauged to realize the GUT. In the gravity dual these flavor branes can be treated in the probe approximation with $N_f\ll N_{\rm CFT}$, ensuring that their backreaction on the AdS$_5\times Y_5$ throat is parametrically small.

In summary, this construction suggests a possible dual 4D interpretation of the warped $SU(5)_{\rm GUT} \times U(1)_{\rm PQ}$ setup: a large-$N_{\rm CFT}$ CFT
with a baryonic-type PQ symmetry, where PQ breaking is mediated by very heavy baryonic operators that need not destabilize the
AdS$_5$ throat, and a single high-quality composite axion that can be identified with the zero mode of $C_5$.


\section{Constraining the Axion Decay Constant}
\label{sec:fa}
\subsection{Perturbativity constraint}\label{sec:fa1}

In a 5D orbifold GUT with a flat extra dimension compactified on $S^{1}/\mathbb{Z}_{2}$, the axion decay constant is determined by the compactification radius $R$ and the gauge coupling constant $g_{5}$ of the bulk gauge theory, i.e. $f_{a}=1/\sqrt{\pi Rg_{5}^{2}}$. Since the logarithmic running of the gauge coupling constants, $g_{i}$, transitions to the linear running (power law) beyond $m_{\rm KK}=R^{-1}$~\cite{Dienes:1998vh,Dienes:1998vg}, maintaining the perturbativity of the unified gauge coupling $g_{\rm GUT}$ at $M_{\rm GUT}$ demands $m_{\rm KK}\gtrsim M_{\rm GUT}$.\footnote{Note that power-law running can actually be used to {\it enhance} the axion mass, an interesting effect~\cite{Gherghetta:2020keg}, that will not be considered in this work.} This, in turn, imposes a non-trivial lower bound on the QCD axion decay constant $f_{a}$. Along this line of reasoning, it was argued in \cite{Benabou:2025kgx} that the QCD axion in a 5D flat orbifold GUT should have a decay constant of order $f_{a}\gtrsim 10^{14}~{\rm GeV}$ with a corresponding upper bound on the mass $m_{a}\lesssim6\times10^{-8}\,{\rm eV}$. 

Given this observation, one may now wonder if the lower bound on $f_{a}$ changes for the QCD axion in a 5D {\it warped} orbifold GUT. As seen in (\ref{eq:fa}), the warped geometry yields an exponentially suppressed decay constant, $f_{a}\propto e^{-\pi k R}$, providing a natural explanation for a parametrically small $f_a$. Moreover, because the gauge couplings continue to evolve only logarithmically, even above the KK threshold $m_{\rm KK}$, perturbativity of the unified gauge coupling $g_{\rm GUT}(M_{\rm GUT})$ does not necessarily require that $m_{\rm KK}\gtrsim M_{\rm GUT}$. Indeed, ``GUT precursors" can appear below the actual unification scale~(see e.g. \cite{Dienes:2002bg}), which in the warped dimension setup corresponds to GUT-charged bound states in the dual CFT. This motivates the question of how far $m_{\rm KK}$ (and hence $f_{a}$) can be lowered in the warped setup while maintaining perturbative unification at $M_{\rm GUT}$.

Although the 4D effective gauge couplings run only logarithmically, perturbativity of $g_{\rm GUT}(M_{\rm GUT})$ can nevertheless be jeopardized in the warped setup because the CFT contribution in (\ref{eq:g4higher}) increases all 4D effective SM gauge couplings at UV scales. Consequently, the key factors determining the perturbativity of $g_{\rm GUT}(M_{\rm GUT})$ are the size of $b_{\rm CFT}$ and the KK scale $m_{\rm KK}$—specifically, how large the former and how small the latter can be before the theory becomes nonperturbative. 
These are related to how large the warping factor $e^{-\pi kR}$ and $N_{\rm CFT}$ are, given $b_{\rm CFT}\simeq N_{\rm CFT}\simeq M_{P}/k$.

\begin{figure}[t]
\centering
\hspace*{-5mm}
\includegraphics[width=0.53\textwidth]{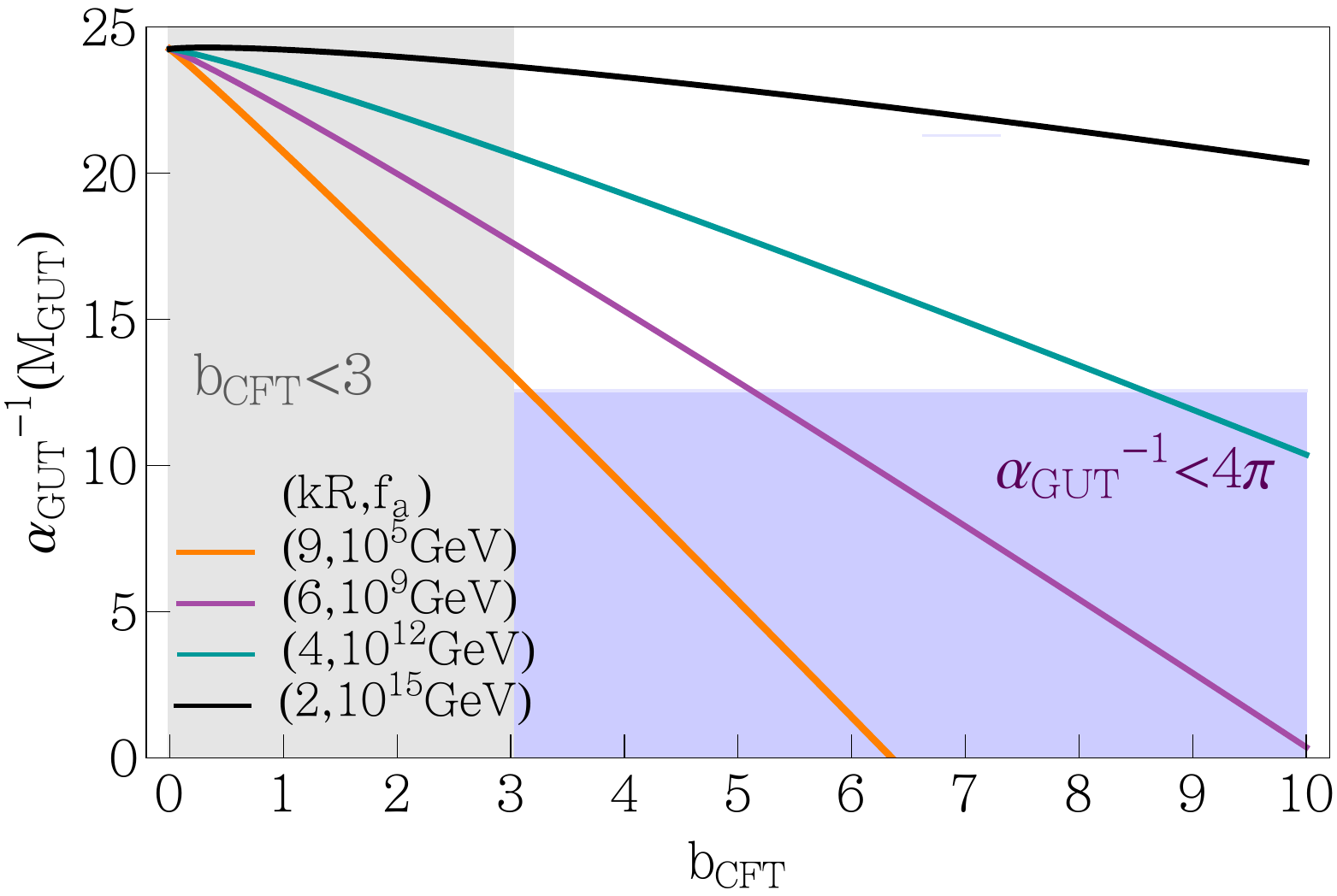}
\caption{The value of $\alpha^{-1}_{\rm GUT}(M_{\rm GUT})$ as a function of $b_{\rm CFT}$. Each {\bf{\color{orange}orange}}, {\bf{\color{violet} purple}}, {\bf{\color{teal}green}} and {{\bf black}} solid line corresponds to $\alpha_{\rm GUT}(M_{\rm GUT})$ for $kR={\color{orange}9}, {\color{violet}6}$ ${\color{teal}4}$ and $2$ respectively. Next to $kR$, we also specify the order of magnitude of the corresponding axion decay constant $f_{a}$. The purple and gray shaded regions indicate, respectively, where $\alpha_{\rm GUT}^{-1}(M_{\rm GUT}) < 4\pi$ (signaling loss of perturbativity) 
and where $b_{\rm CFT} < 3$. Parameter values lying outside these shaded domains remain viable.
}
\vspace*{-1.5mm}
\label{fig:alphaGUT}
\end{figure}

In Fig.~\ref{fig:alphaGUT}, we show $\alpha_{\rm GUT}(M_{\rm GUT})$ as a function of $b_{\rm CFT}$ for several values of $kR$. The orange, purple, green and black lines correspond to $f_{a}=10^{5}~{\rm GeV}, 10^{9}~{\rm GeV}$, $10^{12}~{\rm GeV}$ and $10^{15}~{\rm GeV}$, respectively.  The purple region corresponds to values where $\alpha_{\rm GUT}^{-1}(M_{\rm GUT}) < 4\pi$, signaling the breakdown of perturbativity, while the gray region indicates where $b_{\rm CFT} < 3$. 
Viable parameter choices are those that lie beyond both shaded regions. We see that smaller $f_{a}$ resulting from stronger warping (larger $kR$) leads to larger $\alpha_{\rm GUT}(M_{\rm GUT})$ for a given CFT contribution. This is because stronger warping reduces $m_{\rm KK}$, allowing the CFT contribution to enter at a lower energy scale. For a fixed $m_{\rm KK}$, increasing the CFT contribution (larger $b_{\rm CFT}$) further raises $\alpha_{\rm GUT}(M_{\rm GUT})$. As a result, perturbativity of $g_{\rm GUT}(M_{\rm GUT})$ can be violated when both the warping and CFT contribution are sufficiently large; for example, at $kR=6$, $g_{\rm GUT}$ becomes nonperturbative for $b_{\rm CFT}\gtrsim 5$. Thus, for $kR\leq6$ and $N_{\rm CFT}\simeq b_{\rm CFT}\leq5$, the QCD axion decay constant in a 5D warped orbifold GUT can be as small as $f_{a}=10^{9}\,{\rm GeV}$ without violating the perturbativity of $g_{\rm GUT}(M_{\rm GUT})$. Obviously, smaller values of ($kR,N_{\rm CFT}$) corresponding to $f_{a}\gtrsim 10^{9}{\rm GeV}$ will make it easier to satisfy $\alpha_{\rm GUT}(M_{\rm GUT})\ll1$. Note that for the smallest reasonable choice $b_{\rm CFT}=3$ (consistent with the AdS/CFT correspondence), one has to require 
$kR \gtrsim 9$ in order to prevent the GUT coupling $g_{\rm GUT}(M_{\rm GUT})$ 
from becoming nonperturbative. 
This, in turn, imposes a theoretical lower bound on the axion decay constant, 
$f_{a} \gtrsim 10^{5}\,\mathrm{GeV}$.

\subsection{Small instanton effects}
\label{sec:fa2}

Given that the perturbative unified gauge coupling in the warped orbifold GUT can have much larger values compared to usual 4D supersymmetric GUT ($\alpha_{\rm GUT}(M_{\rm GUT})\simeq 1/24)$, we next examine whether small-instanton effects receive any corresponding enhancement. These small instanton contributions may induce a misaligned term $\delta V(\theta_{a})$ in the QCD axion potential, possibly restricting the allowed values of $f_a$ and $N_{\rm CFT}$.
 
Recall that in a 5D warped orbifold GUT model the presence of the four-fermion, dimension-six operator $QQQL$, where $Q$ and $L$ are SU(2) doublets, is naturally expected. For instance, integrating out the color-triplet Higgs of mass $M_{H_{c}}$ can generate $QQQL$ suppressed by $M_{H_{c}}$. This operator can induce proton decay through $p^{+}\rightarrow\pi^{0}+e^{+}$, and therefore a sufficient suppression of the operator is required. Its presence is natural even when the SU(5) GUT is broken by boundary conditions rather than by a vacuum expectation value of Higgs fields transforming as $\mathbf{24}_{\Phi}$ under SU(5)~\cite{Kawamura:2001ev,Hall:2001pg,Hebecker:2001wq,Goldberger:2002pb}. For instance, when all matter fields---including the Higgs---propagate in the bulk~\cite{Gherghetta:2000qt}, 
the orbifold parities may project out the zero modes of the color-triplet Higgs, 
yet its massive KK excitations can still be integrated out, giving rise 
to a dimension-six operator of the form $\propto QQQL$. In this case, the mass of KK modes of the colored Higgs achieving the largest overlap with zero modes of $Q$ and $L$ will determine the suppression scale of $QQQL$. Even when the SM fermions are localized on the UV brane, higher-dimensional operators of the form $QQQL$ can still be induced by brane-localized or Planck-suppressed interactions. In this case, the suppression scale would be the natural scale of the Planck brane, $M_{P}$.

The presence of higher-dimensional four-fermion operators composed of $SU(2)$ doublets is of particular significance, as such interactions can render the small $SU(2)$ instanton effects dominant~\cite{Kitano:2021fdl,Bedi:2022qrd,Csaki:2023ziz}. Adopting the most conservative standpoint, let us consider the case where the Standard Model fermions are localized on the UV brane. In this setup, one can write the gauge-invariant four-fermion operators as
\be
\mathcal{L}_{4D,{\rm eff}} \supset \frac{c_{ijk\ell}}{M_P^2}\, Q_i Q_j Q_k L_\ell\,,
\label{eq:fourfermion}
\ee
where $c_{ijk\ell}$ are dimensionless, generally complex coefficients, with an implicit sum over generation indices $i,j,k,\ell$. When the operators in (\ref{eq:fourfermion}) are used to close up the zero modes of the $SU(2)$ doublet quarks and leptons emitted from the $SU(2)$ instanton, the instanton density acquires an additional negative power of the instanton size $\rho$, thereby enhancing the ultraviolet contribution to the vacuum-to-vacuum amplitude. This naturally raises the question of whether the small-instanton–induced misalignment in the QCD axion potential, $\Delta\bar{\theta}\simeq \delta V(\theta_{a})/V_{0}(\theta_{a})$, remains sufficiently suppressed to allow $f_a \sim 10^{9}\,{\rm GeV}$ as a viable value where $V_{0}(\theta_{a})\approx\Lambda_{\rm QCD}^{4}=(0.3~{\rm GeV})^{4}$ is the QCD contribution to the axion potential. A naive guess, informed by the possibility that $\alpha_{\rm GUT}(M_{\rm GUT}) \gtrsim 1/24$ once the CFT contributions are included, suggests that the ensuing misaligned term could threaten the quality of the QCD axion, reviving concerns about the consistency of a low-$f_a$ scenario.

As the instanton effect in the slice of AdS$_{5}$ is expected to be equally mimicked by that in the 4D dual theory, let us consider the instanton of the weakly gauged $SU(2)$ symmetry in the 4D dual theory that corresponds to $A_{\mu}$ in (\ref{eq:L4D}). The $SU(N_{\rm CFT})$ gauge theory with “preons" $(\psi,\chi)$ transforming as ($\ytableausetup{textmode, centertableaux, boxsize=0.6em}
\begin{ytableau}
 \\
\end{ytableau} $\,,\,$\overline{\ytableausetup{textmode, centertableaux, boxsize=0.6em}
\begin{ytableau}
 \\
\end{ytableau}} $) under both $SU(N_{\rm CFT})$ and $SU(2)$ forms the CFT sector. For computing the $SU(2)$ small-instanton–induced $\delta V(\theta_{a})$, let us first specify the fermion zero mode legs emitted from the $SU(2)$ instanton. First, there are nine zero modes of quark doublets and three zero modes of lepton doublets which we close up by the operators in (\ref{eq:fourfermion}). For simplicity, we consider the generation diagonal $QQQL$ operators, i.e. $c_{ijk\ell}=c_{i}\delta_{ijk\ell}$. In addition, we have four zero modes of $SU(2)$ gauginos and two zero modes of Higgsinos for which we use the mass insertion. Importantly, there exist also the zero mode legs of the $N_{\rm CFT}$ pairs of preons, which we can close up with bilinear condensates formed from the confinement of $SU(N_{\rm CFT})$ at $q\simeq \Lambda_{N_{\rm CFT}}$. This amounts to the instanton density $(\rho\Lambda_{N_{\rm CFT}})^{N_{\rm CFT}}$ ~\cite{Gherghetta:2021jnn}.

\begin{figure}[t]
\centering
\hspace*{-5mm}
\includegraphics[width=0.4\textwidth]{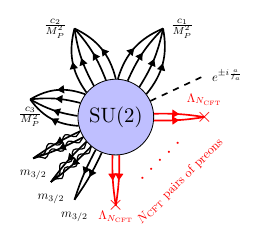}
\caption{An $SU(2)$ instanton diagram assuming small instantons of size $\rho\simeq M_{\rm GUT}^{-1}$. The zero mode lines of the $SU(2)$ doublet quarks and leptons are closed up by the $c_{i}\frac{QQQL}{M_{P}^{2}}$ operators with $i=1,2,3$ the generation index. The two Higgsino zero mode solid lines and the four $SU(2)$ gaugino zero mode wavy lines are closed up by the mass insertions with the gravitino mass, $m_{3/2}$, parametrizing the supersymmetry-breaking soft masses.  The red lines correspond to the preon zero mode legs and there are $N_{\rm CFT}$ pairs of closed legs.
}
\vspace*{-1.5mm}
\label{fig:su2instanton}
\end{figure}

Closing the fermion zero modes emitted from the $SU(2)$ instanton, as described above, yields the diagram shown in Fig.~\ref{fig:su2instanton}, which represents the dominant misaligned contribution to the QCD axion potential arising from small $SU(2)$ instantons. In this analysis, the soft SUSY-breaking masses are parametrized by the gravitino mass $m_{3/2}$. The red lines in Fig.~\ref{fig:su2instanton} denote the preon zero-mode legs, with $N_{\rm CFT}$ pairs forming closed loops. Crucially, owing to the generic complex phases of the coefficients $c_i$, the $SU(2)$ instanton diagram generically induces a misaligned contribution to the QCD axion potential relative to the standard QCD-confinement-induced potential $V_0(\theta_a)$.\footnote{Analogously, there can also be gauge-invariant four-quark operators of the form $u_{L}\,\bar{u}_{R}\,d_{L}\,\bar{d}_{R}/M_{P}^{2}$, where the Weyl fields $u_{L,R}$ and $d_{L,R}$ transform in the ${\bf 3}$ of $SU(3)$, and $\bar{u}_{R}$ denotes the Hermitian conjugate of $u_{R}$. These operators can be used to close up the twelve quark zero modes generated by a small $SU(3)$ instanton, thereby rendering the corresponding UV-instanton contribution dominant. Including the six $SU(3)$ gluino zero modes, and assuming Wilson coefficients of these four-quark operators comparable in magnitude to $c_{ijk\ell}$ in (\ref{eq:fourfermion}), the resulting GUT-scale $SU(3)$ instanton contribution can be of the same order as the effect depicted in Fig.~\ref{fig:su2instanton}.
}

In order for QCD axion to explain the null observation of the neutron electric dipole moment thus far, $\delta V(\theta_{a})$ from Fig.~\ref{fig:su2instanton} needs to be constrained by~\cite{Baker:2006ts}
\be
\Delta\bar{\theta}\simeq\frac{\delta V(\theta_{a})}{V_{0}(\theta_{a})}<10^{-10}\,.
\label{eq:deltatheta}
\ee
Using instanton NDA~\cite{Csaki:2023ziz}, we find $\delta V(\theta_{a})$ contributed by $SU(2)$ instantons of size $\rho=M_{\rm GUT}^{-1}$ from Fig.~\ref{fig:su2instanton} to be
\ba
\delta V(\theta_{a})&\simeq&\left(\frac{2\pi}{\alpha_{2}(M_{\rm GUT})}\right)^{4}\frac{\Pi_{j=1}^{3}|c_{j}|}{(4\pi)^{6}}\left(\frac{M_{\rm GUT}}{M_{P}}\right)^{6}\cr\cr
&\times&\left(\frac{m_{3/2}}{M_{\rm GUT}}\right)^{3}\left(\frac{\Lambda_{N_{\rm CFT}}}{M_{\rm GUT}}\right)^{N_{\rm CFT}}\cr\cr
&\times&M_{\rm GUT}^{4}e^{-\frac{2\pi}{\alpha_{2}(M_{\rm GUT})}}\cos\left[\theta_{a}+\sum_{j=1}^{3}\delta_{j}\right]\,,\nonumber\\
\label{eq:deltaV}
\ea
where $c_{j}=|c_{j}|e^{i\delta_{j}}\in\mathbb{C}$ and $\alpha_{2}$ is given in Eq.~(\ref{eq:g4higher}). For $m_{3/2}=100\,{\rm TeV}$ and $c_{i}=\mathcal{O}(1)$, on comparing $\delta V(\theta_{a})$ in (\ref{eq:deltaV}) with $\Lambda_{\rm QCD}^{4} \simeq (0.3\,{\rm GeV})^{4}$, we find numerically that $\Delta\bar{\theta} = \mathcal{O}(10^{-44}) \ll 10^{-10}$, thereby dispelling the naive concern that a unified coupling $\alpha_{\rm GUT}$ greater than 4D SUSY GUT prediction could amplify the small-instanton effect and jeopardize the axion solution. This result is surprisingly insensitive to values of $N_{\rm CFT}$ as long as $\Lambda_{N_{\rm CFT}}\simeq m_{\rm KK}$ and $N_{\rm CFT}\simeq b_{\rm CFT}$. But then, how can $\delta V(\theta_{a})$ be sufficiently suppressed even for $\alpha_{\rm GUT}$ close to 1?

This surprising “chiral suppression" effect was first observed in \cite{Gherghetta:2021jnn} where the vacuum-to-vacuum amplitude was computed by expanding the 5D action around a 5D instanton solution yielding an extra suppression due to a large effective action. This effect can be understood in the dual 4D theory by assuming that the CFT necessarily contains fermionic preons charged under the SM gauge group which must then be included in the small instanton diagram. When the zero modes of the preons are closed up, an additional factor in the instanton density cancels their contribution to the running of the gauge coupling in (\ref{eq:g4higher}). In (\ref{eq:deltaV}), this cancellation manifests as
\ba
&&\left(\frac{\Lambda_{N_{\rm CFT}}}{M_{\rm GUT}}\right)^{N_{\rm CFT}}e^{-\frac{2\pi}{\alpha_{2}(M_{\rm GUT})}}\cr\cr
&&\simeq e^{-\frac{2\pi}{\alpha_{2}(M_{\rm GUT})}+b_{\rm CFT}\ln\left(\frac{\Lambda_{N_{\rm CFT}}}{M_{\rm GUT}}\right)}=e^{-\frac{2\pi}{\alpha^{(0)}_{2}(M_{\rm GUT})}}\,
\ea
where $\alpha^{(0)}_{2}(M_{\rm GUT})\simeq 1/24$ is the running coupling due solely to the zero modes, evaluated at the GUT scale. This cancellation matches with the 5D calculation where a larger effective action was obtained.

This observation indicates that when the extra-dimensional axion framework is embedded within a 5D warped orbifold GUT, the resulting QCD axion decay constant can naturally reside within the conventional window, $10^{9}\,{\rm GeV} \lesssim f_{a} \lesssim 10^{12}\,{\rm GeV}$, without violating perturbativity of the unified gauge coupling, $\alpha_{\rm GUT}(M_{\rm GUT}) < 1$, or spoiling the axion quality. Moreover, perturbative gauge coupling unification can still be successfully achieved for $kR \lesssim 6$ and $b_{\rm CFT} \lesssim 5$. For larger values of $b_{\rm CFT}$—as might be expected in the large-$N_{\rm CFT}$ limit—the perturbativity condition $\alpha_{\rm GUT}(M_{\rm GUT}) < 1$ can nonetheless be maintained provided that $kR$ is correspondingly smaller, which in turn yields a larger value of $f_{a}$.


\section{Conclusion}\label{sec:conclusion}

In this paper, we have examined how the QCD axion decay constant is constrained when the axion originates from a 5D $U(1)_{C}$ gauge field propagating in a slice of AdS$_5$, with the SM gauge fields embedded in a warped orbifold grand unified theory. Our analysis makes crucial use of the fact that, unlike in flat extra dimensions where the gauge coupling exhibits a power-law running above the KK threshold, the effective 4D gauge coupling in the warped geometry continues to evolve only logarithmically, even for momenta above $m_{\rm KK}$. This slower running substantially alters the relation between the axion decay constant $f_{a}$, the compactification scale, and the scale of gauge coupling unification.

We have shown that maintaining perturbative unification at $M_{\rm GUT}$ permits a substantially broader range of $f_{a}$ than in the flat scenario, potentially spanning the canonical QCD axion window, $10^{9}\,{\rm GeV} \lesssim f_{a} \lesssim 10^{12}\,{\rm GeV}$. The viable ranges of $kR$ and $b_{\rm CFT}$ are depicted in Fig.~\ref{fig:alphaGUT}. Although one might worry that the axion quality could be compromised for low $f_{a}$—since small instanton effects can be enhanced by unavoidable four-fermion operators such as those in (\ref{eq:fourfermion}), leading to misaligned contributions to the axion potential—we have demonstrated that the axion quality remains robust due to chiral suppression. Consequently, the warped construction offers a phenomenological advantage over the flat case, as lower values of $f_{a}$ enhance the experimental accessibility of axion searches at low energies. 

More broadly, the 5D warped setup can be viewed as an effective description of warped flux compactifications in string theory, in which axions arise from higher-dimensional (Ramond-Ramond sector) higher-form gauge fields. Constructing an explicit UV completion along these lines remains an interesting direction for future work.

Furthermore, we have proposed that the 5D warped model admits a dual 4D interpretation, in which the PQ symmetry could be identified with a baryon symmetry, $U(1)_B$, that is spontaneously broken on the baryonic branch 
of the strongly-coupled dual theory (e.g. the Klebanov-Witten theory).
The QCD axion is then the pNGB associated with a baryon operator, while the $SU(5)$ grand unified group (containing the Standard Model gauge group) weakly gauges an $SU(N_f)$ flavor subgroup of the dual 4D gauge theory. Bound states appear at the confinement scale (dual to IR Kaluza-Klein modes), above which the running of the SM gauge couplings is modified. Since only baryon operators carry PQ charge (or baryon number), PQ-violating operators induced by gravity are highly suppressed at large $N_{\rm CFT}$. This feature is dual to the exceptional high-quality axion arising from the $C_5$ zero-mode of the bulk U(1) gauge field in the slice of AdS$_5$.

In the present analysis, we have adopted a conservative estimate of the QCD axion potential misalignment induced by small $SU(2)$ instantons by assuming that all SM fermions are confined to the UV brane, leading to strong suppression of higher-dimensional operators such as $QQQL$. It would, however, be worthwhile to revisit this estimate in a more realistic setup with bulk Standard Model fermions, whose localization profiles can simultaneously account for the observed Yukawa hierarchies. We leave a detailed investigation of this possibility for future work.

Our results highlight how warped geometry provides a natural framework connecting the QCD axion, gauge coupling unification, and holography. In particular, our construction gives a concrete example in which an “extra-dimensional axion”—realized as the Wilson-line zero mode of a 5D gauge field—and a “field-theory axion”—emerging as a pNGB of a spontaneously broken global symmetry in the dual 4D description—are simply two complementary descriptions of the same QCD axion. The persistence of logarithmic running above the KK scale, despite an infinite tower of resonances, plays a central role in reconciling perturbative unification with phenomenologically viable axion decay constants. We expect that further exploration of this overlap between geometric and dynamical axion realizations will yield new insights into the origin and robustness of axions beyond the Standard Model.

\medskip\noindent\textit{Acknowledgments\,---\,}%
We thank Kiwoon Choi, Katherine Fraser, Alex Pomarol, Matt Reece, Chang Sub Shin and Matt Strassler for helpful discussions.
This work was supported by the Department of Energy under Grant No. DE-SC0011842 at the University of Minnesota.

\appendix

\section{$C_{5}$ profile}
\label{app:c5profile}

In this Appendix, we review the calculation of the zero mode profile $C_{5}$ in the 5D warped extra dimension.

Consider the 5D action of a $U(1)_{C}$ gauge theory with gauge kinetic term
\begin{equation}
S_{\text{kin}} = \frac{1}{4g_{5C}^{2}}
\int d^{4}x\,dy\,\sqrt{-g}\,C_{MN}C^{MN}\,,
\label{eq:5Dkinetic}
\end{equation}
and the following gauge fixing term to remove mixing between $C_{\mu}$ and $C_{5}$
\begin{equation}
S_{\text{GF}}
= -\int d^{4}x\,dy\,
\frac{e^{-2ky}}{2\xi g_{5C}^{2}}
\left(\partial_{\mu}C^{\mu}
+ \xi\,\partial_{5}C^{5}\right)^{2}\,,
\label{eq:SGF}
\end{equation}
where $\xi$ is a gauge fixing parameter. After the mixing is canceled, the remaining $C_{5}$ part of the quadratic action reads
\begin{equation}
S_{5D}
\supset \frac{1}{2g_{5C}^{2}}
\int d^{4}x\,dy\,e^{-2ky}
\left[
(\partial_{\mu}C_{5})^{2}
- \xi\,(\partial_{5}C_{5})^{2}
\right]\,.
\label{eq:woGF}
\end{equation}
Using (\ref{eq:woGF}), the $C_{5}$ equation of motion is then
\begin{equation}
-\,e^{-2ky}\Box_{4}C_{5}
+ \xi\,\partial_{y}\!\left(e^{-2ky}\partial_{5}C_{5}\right)=0\,,
\label{eq:eomC5}
\end{equation}
where $\Box_{4}$ is 4D d'Alembertian operator. The second term on the left hand side of (\ref{eq:eomC5}) is relevant for the $C_{5}$ zero mode profile. Thus, from 
\begin{equation}
\partial_{y}\!\left(e^{-2ky}\partial_{5}C_{5}\right)=0\,,
\end{equation}
we obtain the general form of the $C_{5}$ zero mode profile
\begin{equation}
C_{5}(y)=\tilde{A} + \tilde{B}\,e^{2ky}\,,
\end{equation}
where $\tilde{A}$ and $\tilde{B}$ are integration constants. Imposing the boundary condition, $C_{5}(-y)=C_{5}(y)$, on $C_{5}$ at each orbifold fixed point then enforces $\tilde B=0$, which leaves the $y$-independent zero mode of $C_{5}$~\cite{Choi:2003wr}.

\section{4D Axion Decay Constant}
\label{app:fa}

In this Appendix, we present the calculation of the axion decay constant.

The $C_{M}$ equation of motion from the kinetic term in (\ref{eq:S5D}) gives
\be
\partial_{5}\!\left[e^{-2ky}\,(\partial_\mu C_5 - \partial_5 C_\mu)\right]=0\,.
\label{eq:eomc5}
\ee 
Using the result from Appendix~\ref{app:c5profile} that $C_{5}$ is $y$-independent, (\ref{eq:eomc5}) can be rewritten as 
\be
\partial_5 C_\mu(x,y) = \partial_\mu C_5(x) - D_\mu(x)e^{2ky}\,,
\label{eq:d5Cxy}
\ee
where $D_\mu(x)$ is an integration function. Given that $C_{\mu}$ satisfies the orbifold boundary conditions $C_\mu(0)=C_\mu(\pi R)=0$, integrating (\ref{eq:d5Cxy}) over $y$ from $0$ to $\pi R$, one obtains
\be
D_\mu(x) = \frac{2\pi kR}{e^{2\pi kR}-1}\,\partial_\mu C_5(x)\,,
\ee
and thus
\be
C_{\mu5}(x,y)
= \frac{2\pi kR\,e^{2ky}}{e^{2\pi kR}-1}\,\partial_\mu C_5(x)\,.
\label{eq:Cmu5-solution}
\ee
Substituting \eqref{eq:Cmu5-solution} into the 5D kinetic term gives
\be
S_{5D}
\supset -\int d^4x\,dy\,
\frac{e^{2ky}}{4g_{5C}^{2}}\left(\frac{2\pi kR}{e^{2\pi kR}-1}\right)^{2}
\,\partial_\mu C_5\,\partial^\mu C_5\,.
\label{eq:S4Dkind5x}
\ee
From (\ref{eq:S4Dkind5x}), we infer that the profile of $C_{5}$ with respect to the 5D flat metric is proportional to $e^{ky}$ and thus localized towards the IR brane. Integrating over 
$y$, 
we obtain the 4D kinetic term for the zero mode $C_5(x)$:
\be
S_{4D,\rm kin}
=\int d^4x\,
\frac{\pi^2 k R^2}{2g_{5C}^2}\frac{1}{e^{2\pi kR}-1}
\,\partial_\mu C_5\,\partial^\mu C_5\,.
\label{eq:S4Dkin}
\ee
In addition, the 5D Chern--Simons term in (\ref{eq:S5D}) reduces to 4D term
\be
S_{4D,\rm CS}
=\int d^4x\, \frac{\pi R c_{5}}{16\pi^{2}}\, C_5(x)\,G^{a}_{\mu\nu}\tilde G^{a\,\mu\nu}\,,
\label{eq:S4DCS}
\ee
where $G^{a}_{\mu\nu}$ is the field strength of $SU(5)_{\rm GUT}$ gauge field. To canonically normalize the 4D axion field $a(x)$, we define 
\begin{equation}
a(x)=\sqrt{2K}\,C_5(x),
\qquad
K=\frac{\pi^2 k R^2}{2g_{5C}^2}\frac{1}{e^{2\pi kR}-1}.
\label{eq:defineaxion}
\end{equation}
The 4D effective Lagrangian then takes the form
\begin{equation}
\mathcal{L}_{4D}
=-\frac{1}{4g_{\rm GUT}^{2}}G^{a}_{\mu\nu}G^{a\,\mu\nu}+\frac{1}{2}(\partial_\mu a)^2
+\frac{c_{5}}{32\pi^2}\frac{a}{f_a}G^{a}_{\mu\nu}\tilde G^{a\,\mu\nu}\,,
\label{eq:4DSeff}
\end{equation}
where we have used the relation $\pi R/g_{5}^{2}=g_{\rm GUT}^{-2}$ and assumed a constant zero-mode profile of the $SU(5)$ gauge field.

Now, matching (\ref{eq:4DSeff}) to (\ref{eq:S4Dkin}) and (\ref{eq:S4DCS}) yields~\cite{Choi:2003wr}
\begin{equation}
f_a = \frac{\sqrt{2K}}{2\pi R}
= \sqrt{\frac{k}{4g_{5C}^2}\frac{1}{e^{2\pi kR}-1}}.
\label{eq:fa-final}
\end{equation}
Finally, one can check that (\ref{eq:defineaxion}) and (\ref{eq:fa-final}) are consistent with the identification in (\ref{eq:thetaax}).


\bibliographystyle{JHEP}
\bibliography{arxiv_1}


\end{document}